\documentclass[twocolumn,prb,showpacs]{revtex4}
\usepackage{graphicx}
\usepackage{textcomp}
\usepackage{bm}
\newcommand\Rb{$^{87}$Rb}
\newcommand\qtd{quasi two-dimensional}
\newcommand\afm{antiferromagnet}

\newcommand\RbFe{$\rm RbFe(MoO_4)_2$}
\newcommand\PRB[3]{Phys. Rev. B {\bf {#1}}, {#2} ({#3})}

\newcommand\JPCM[3]{J. Phys. Condens. Matter {\bf {#1}}, {#2} ({#3})}
\newcommand\JPSJ[3]{J. Phys. Soc. Jpn. {\bf {#1}}, {#2} ({#3})}

\begin{document}
\title{Magnetic phase diagram, critical behaviour and 2D-3D crossover in a triangular lattice antiferromagnet RbFe(MoO$_4$)$_2$}
\author{L.~E.~Svistov, A.~I.~Smirnov, L.~A.~Prozorova}
\affiliation{P.~L.~Kapitza  Institute for  Physical  Problems RAS, 117334
Moscow, Russia}
\author{O.~A.~Petrenko}
\affiliation{Department of Physics, University of Warwick, Coventry, CV4
7AL, UK}
\author{ A.~Micheler, N. B\"uttgen}
\affiliation{Experimentalphysik V, Center for Electronic Correlations and
Magnetism, University of Augsburg, D-86135 Augsburg, Germany}
\author{A.~Ya.~Shapiro, L.~N.~Demianets}
\affiliation{A.~V.~Shubnikov  Institute for  Crystallography  RAS, 117333
Moscow, Russia}
\date{\today}
\begin{abstract}
We have studied the magnetic and thermodynamic properties as well as the
NMR spectra of the \qtd\ Heisenberg \afm\  \RbFe. The observed
temperature dependence of the order parameter, the critical indices and
the overall magnetic $H-T$ phase diagram are all in a good agreement with
the theoretical predictions for a 2D XY model. The temperature dependence
of the specific heat at low temperature demonstrates a crossover from a
$T^2$ law characteristic of a two-dimensional \afm\ to a
three-dimensional $T^3$ law.
\end{abstract}
\pacs{75.50.Ee; 76.60-k.}

\maketitle
\section{Introduction}
The problem of finding the ground state of a two-dimensional triangular
lattice \afm\ (TLAM) is of particular interest due to the possibility of
finding solutions with unconventional magnetic order, which are
influenced significantly both by frustration and by zero-point
fluctuations. As the magnetic exchange interactions between the ions
located on a regular two-dimensional triangular lattice are frustrated, a
usual Neel-type ground state with anti-parallel alignment of the nearest
neighbour moments is hindered by the geometry. In the molecular field
approximation, the minimum of energy is reached for a planar spin
configuration with the magnetisation of the three sublattices arranged at
120$^\circ$ to one other. Such a spin configuration has been found in
numerous compounds with a stacked triangular lattice \cite{Collins},
regardless of the dimensionality of the magnetic interactions, which can
have a pronounced 1D or 2D character. A very recent report suggests that
instead of a long range structure, a spin-liquid state with a relatively
short correlation length may be stabilised in some compounds
\cite{Nakatsuji}.

In an applied magnetic field, the mean-field ground state of the 2D
triangular \afm\ remains highly degenerate, as the overall energy is
defined only by the total spin of the three sublattices, while there is
an infinitely large number of different states with the same total spin
\cite{Kawamura,Landau}. Indeed, any possible value of the total spin can
be represented as an umbrella-type structure of equally tilted magnetic
moments with identical transverse components, or, alternatively, as a
variety of both planar and nonplanar structures with variable angles
between the magnetic moments and the direction of the field.

This degeneracy is removed by thermal and quantum fluctuations, which
select a symmetric planar structure for a purely isotropic (Heisenberg)
case,  for an easy-axis type of anisotropy and also for an easy-plane
type of anisotropy if the field is applied in the basal plane
\cite{Chubukov,Gekht,Rastelli,Korshunov}. If the field is applied
perpendicular to the easy-plane, the mean-field approach yields an
umbrella-type structure, which is non-degenerate with respect to the
mutual orientation of magnetic moments. For the planar structures,
fluctuation analysis reveals a characteristic feature of the
magnetisation curve $M(H)$. Namely, a collinear magnetic structure with
nearly one third of the saturation magnetisation ($M=M_s/3$) that is
stabilised by the fluctuations in the vicinity of the field $H_s/3$,
where $H_s$ is the saturation field. The stabilisation of this structure
is seen as a plateau on the $M(H)$ curve, while in the molecular field
approximation the magnetisation curve has a continuous derivative. This
structure is referred to as ``two spins up, one down" or UUD.

\begin{figure}
\includegraphics[width=0.9\columnwidth]{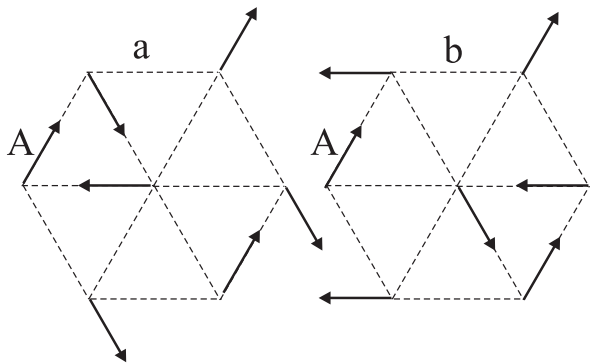}
\caption{Two different magnetic structures on a hexagonal lattice
possessing the same exchange energy. The direction of magnetic moment at
point A is identical for both structures.}
 \label{chir}
\end{figure}

Apart from this, TLAMs are distinguished from conventional collinear ferro
and \afm s by the fact that for some phases, called {\it chiral} phases,
there are still two ways to arrange the rest of the structure even when
the direction of the spin on a particular apex of the triangle is fixed.
Fig.  \ref{chir} shows two magnetic structures, a and b, which share the
alignment of the spin in the position A and possess the same exchange
energy, but differ by the direction of rotation of the spins when
translated to the nearest apex. Chiral degeneracy influences considerably
the nature of the phase transition to an ordered state
\cite{Miyashita,Kawamura}. Within a magnetic $H-T$ phase diagram for a
single TLAM, both chiral and nonchiral structures are possible, therefore
the second order phase transitions with different critical indices should
be expected for the different values of an applied magnetic field.

An especially interesting point to consider is a transition from a
paramagnetic phase to a UUD structure occurring in a constant field on
lowering the temperature. This is a transition into a collinear phase
with an uncompensated magnetic moment. Such structures with uncompensated
magnetic moment are characteristic of ferrimagnets, therefore for a
description of this collinear phase on a triangular lattice a
ferrimagnetic order parameter is used \cite{Miyashita}. In ordinary
ferrimagnets, however, the magnetic ions are located on non-equivalent
crystallographic positions, which leads to the appearance of
non-equivalent magnetic sublattices and an uncompensated moment. In these
conventional ferrimagnets an ordering transition in an applied magnetic
field is absent, as an induced moment exists at all temperatures. In the
case of a 2D TLAM, however, all the magnetic ions are in equivalent
positions, while the ordering into a UUD structure is accompanied by a
tripling of the unit cell area, which causes a transition of a specific
type \cite{Landau}.

The majority of TLAMs have quasi one-dimensional character. These systems
have been known and studied for some time now\cite{Collins}. 2D TLAMs
demonstrating the properties described above are less common and have
only recently become the subject of intense investigations. \RbFe\ offers
a rare opportunity to study a model system, which resembles closely an
ideal 2D TLAM. The crystal structure of \RbFe\ described by the space
group $D_{3d}^3$ consists of layers of magnetic Fe$^{3+}$ ions separated
by (MoO$_4$)$^{2-}$ groups and Rb$^+$ ions. The $C^3$ axis is
perpendicular to the layers. The magnetic Fe$^{3+}$ ions form an
equilateral triangular lattice within the layers, whereas along the $C^3$
axis they are interspaced by the Rb$^+$ ions. Such a layered structure
(depicted in References \onlinecite{Klevtsov} and  \onlinecite{Klimin})
ensures the magnetic two-dimensionality of \RbFe \cite{Klevtsov,Svistov}.

Relatively weak exchange interactions in \RbFe\ allow one to reach the
saturation field experimentally, while the large spin ($S=5/2$) of the
magnetic Fe$^{3+}$ ions justify a quasi classical description of the
magnetic system. Previous experiments on powder and single crystal
samples of \RbFe\ established the presence of long range magnetic order
below $T_N=3.8$~K and also demonstrated the stabilisation of the
collinear structure around $M_s/3$ \cite{Inami,Svistov}. Specific heat
measurements \cite{Jame} gave estimates of the magnetic entropy change
during the ordering process. The dynamics of low-frequency excitations
probed by ESR measurements \cite{Svistov} gave an estimate for the ratio
of the intraplane to interplane exchange interactions as 25 and were also
used to evaluate the magnetic anisotropy. By measuring the field and
temperature dependence of the magnetisation  \cite{Svistov} and the
specific heat \cite{Jame}, the magnetic $H-T$ phase diagram was mapped
out, the overall shape of which agreed well with the theoretical
predictions for a 2D TLAM \cite{Korshunov,Miyashita,Kawamura,Landau}.

Neutron diffraction measurements performed in zero field
\cite{Broholm2002} have found an ordered structure with a tripled in-plane
period, which corresponds to a 120-degree configuration within the layers
and an incommensurate ordering along the $C^3$ axis.  NMR spectra of the
\Rb\ ions located between the magnetic Fe$^{3+}$ ions from neighbouring
planes were reported in ref. \onlinecite{NMR}. These spectra have allowed
identification of various low-temperature phases:  a structure with
incommensurate modulation along the $C^3$ axis in low fields, a tilted
commensurate structure and a UUD structure in higher fields.

This paper presents the results of a detailed investigation of the
magnetic phase diagram and critical behaviour in the different phases of
\RbFe\ obtained by thermodynamic measurements. It also reports on the
temperature dependence of the order parameter obtained by the \Rb\ NMR
spectroscopy. The magnetic properties, phase diagram and critical indices
demonstrate good quantitative agreement with the theoretical predictions
for a classical 2D XY-model.

\section{Experimental details}

Single-crystal samples of \RbFe\ were synthesised by means of spontaneous
crystallisation as described in Ref. \onlinecite{Klevtsov}. The samples
were thin (0.5 mm), nearly hexagonal plates with lengths along the edges
up to 6 mm. The $C^3$-axis was found to be perpendicular to the surface of
the plates for all samples. The results reported in this paper were
obtained using samples produced by a significantly slower cooling of the
melt compared to the samples used in Ref.~\onlinecite{Svistov,Jame}
(1~K/hour {\it vs} 3~K/hour). This has resulted in samples with increased
thickness and linear dimensions.

The magnetic moment was measured in a field of up to 12~T using an Oxford
Instruments vibrating sample magnetometer. The heat-capacity measurements
were performed by the standard heat-pulse method in the temperature range
0.4 to 40~K in a field of up to 9~T using a Quantum Design PPMS
calorimeter equipped with a $^3$He option. NMR spectra of the \Rb\ ions
($I=3/2, \gamma =13.9312$~MHz/T) were recorded on the home-made
spectrometer covering the range 35 to 114~MHz. The spin-echo signal was
observed at constant frequency on sweeping the magnetic field in the
range 2.5-9~T. Radio frequency pulses of 5~$\mu$s and 10~$\mu$s were
separated by the time interval $\tau_D=50$~$\mu$s.

\section{Experimental results}

\subsection{Magnetisation and specific heat measurements}
\label{subsec_Magnet}
\begin{figure}
\includegraphics[width=0.95\columnwidth]{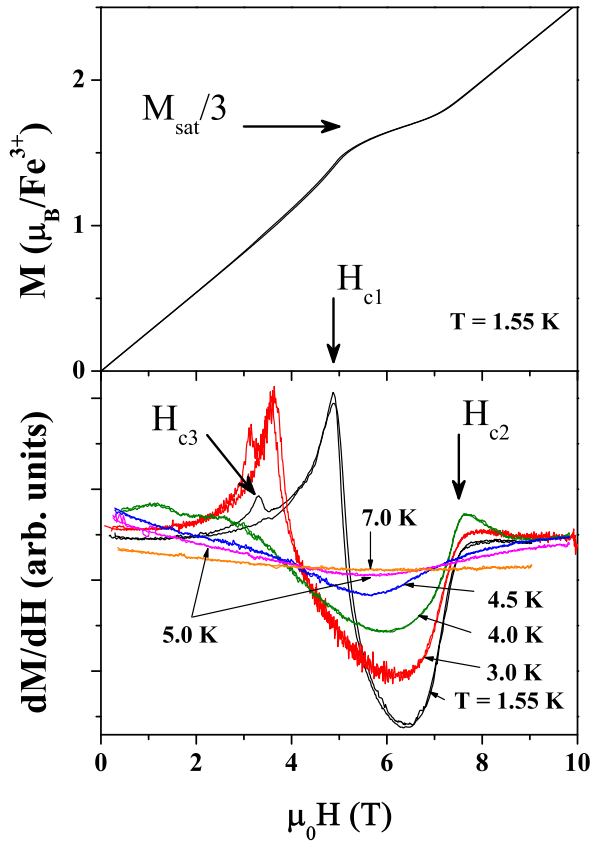}
\caption{(Color online) Magnetisation curve versus field of a \RbFe\
single crystal at $T=1.55$~K (top panel) and the derivative,
$\frac{dM}{dH}$, taken at different temperatures (bottom panel).
               The field is applied perpendicular to the $C^3$-axis.}
\label{fig:MvsH}
\end{figure}

\begin{figure}[tb]
\includegraphics[width=0.95\columnwidth]{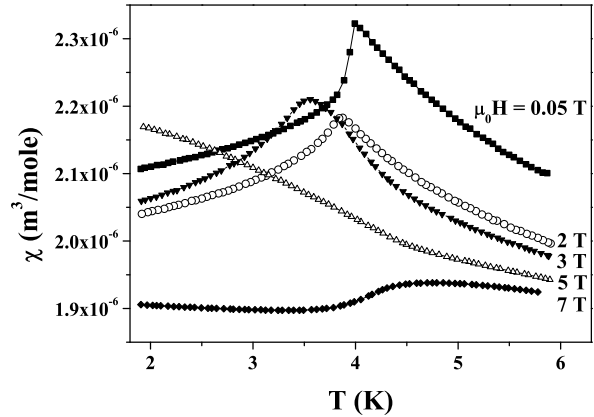}
\caption{Temperature dependence of the magnetic susceptibility of a
single crystal of \RbFe, ${\bf H} \perp C^3$ measured in different
magnetic fields.} \label{fig:MvsT}
\end{figure}

Field and temperature dependence of the magnetic moment of \RbFe\ are
shown in Figs. \ref{fig:MvsH} and \ref{fig:MvsT}, respectively. The $M(H)$
and especially the $dM/dH$ curves show the $M_s/3$ plateau in the field
range $(H_{c1},H_{c2})$ as well as an additional phase transition at
$H_{c3}$ accompanied by a weak hysteresis. The obtained data are very
similar to the  $M(H)$ curves reported previously \cite{Svistov}, apart
from slightly reduced hysteresis effects around the $H_{c3}$ field and
also a small (about 0.1~K) increase in the ordering temperature compared
to the value of $T_N$ obtained on the batch of the samples studied
beforehand.\cite{Svistov,Jame,Broholm2002} For this new sample batch, the
relative amplitude of the susceptibility peak at $T_N$ has increased by a
factor of two for small applied fields, which usually indicates an
improved sample quality. The anomalies, singularities and other special
points, which are seen in the curves in Figs. \ref{fig:MvsH} and
\ref{fig:MvsT} and which correspond to various phase transitions are used
to locate the positions of the phase boundaries, which are later
summarised in the magnetic phase diagram (Fig. \ref{PhaseDiagr}).

\begin{figure}[tb]
\includegraphics[width=0.95\columnwidth]{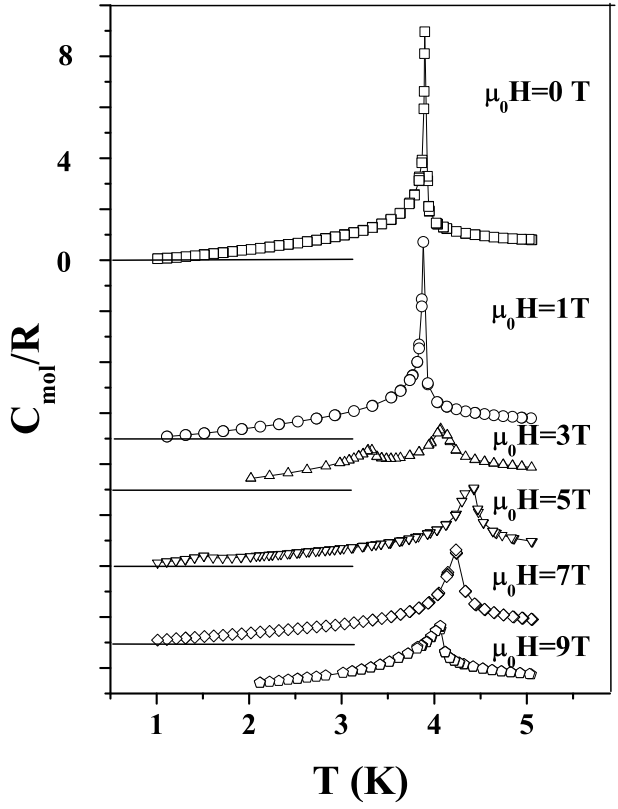}
\caption{Temperature dependence of the specific heat measured in \RbFe,
${\bf H} \perp C^3$ measured in different magnetic fields.
                The curves are offset along the vertical axis.}
 \label{fig:CTip1}
\end{figure}

Fig. \ref{fig:CTip1} shows the temperature dependence of the specific
heat measured in \RbFe\ for various values of a magnetic field applied in
the basal plane of the crystal. The $C(T)$ curves demonstrate sharp
anomalies corresponding to the phase transition from the high temperature
paramagnetic phase to the low temperature magnetically ordered phase in
every field from 0 to 9~T. For the $C(T)$ curves measured in the field
interval 2.5 to 5.5~T, additional peaks are also observed when the
external field coincides with the value of $H_{c1}$.

\begin{figure}
\includegraphics[width=0.99\columnwidth]{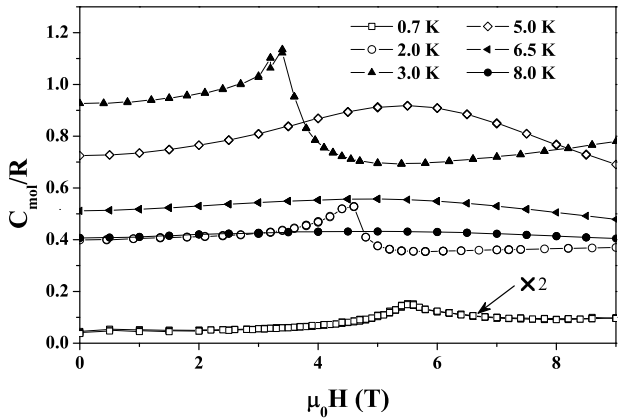}
\caption{Field dependence of the specific heat of \RbFe\ measured for
${\bf H} \perp C^3$ for various temperatures.}
 \label{fig:CHip}
\end{figure}

Fig. \ref{fig:CHip} shows the field dependence of the specific heat
measured at different  temperatures (${\bf H} \perp C^3$). For
temperatures below 4~K, sharp peaks associated with the start of the
plateau at $H_{c1}$ are clearly seen on the $C(H)$ curves, while the
phase transitions at $H_{c2}$ or $H_{c3}$ could not be detected here. For
temperatures above 4.5~K, a wide and round maximum in $C(H)$ develops
around a field of 5~T. On further heating, this maximum reduces in
amplitude and becomes undetectable above 8~K. The observed peak positions
were plotted on the phase diagram (Fig. \ref{PhaseDiagr}) together with
the data obtained from the magnetisation measurements.

As can be seen from Fig.  \ref{fig:CTip1}, the application of a weak
magnetic field reduces marginally the ordering temperature. However, when
the field is strong enough to stabilise a collinear phase, the ordering
temperature develops a non-monotonic dependence with applied field. It
rises with field for $H>H_{c1}$ reaching a maximum of 4.5~K at $H=5$~T.
This maximum value exceeds significantly the zero-field ordering
temperature of 3.9~K.

\begin{figure}
\includegraphics[width=0.95\columnwidth]{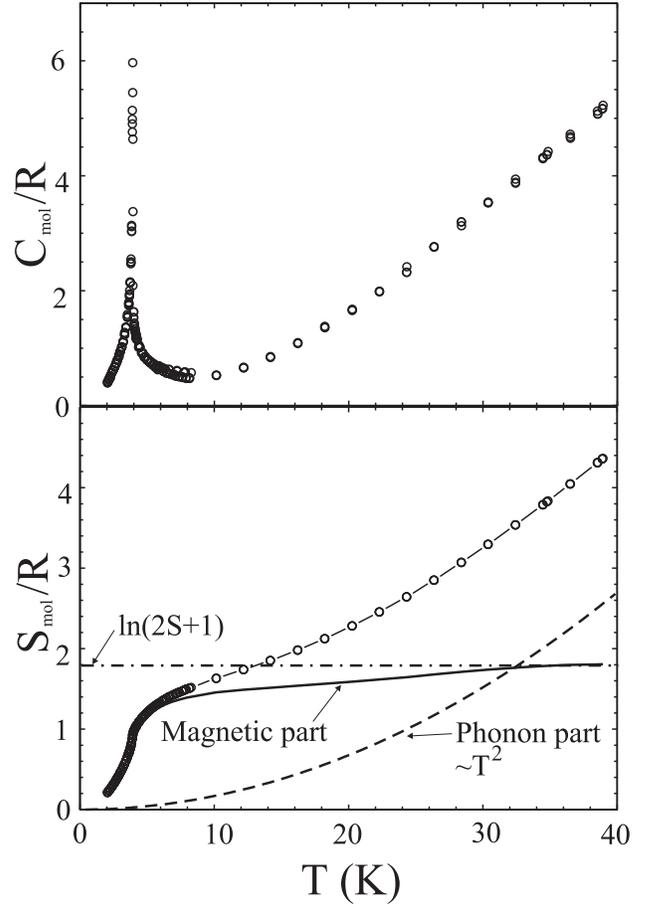}
\caption{Temperature dependence of the specific heat (top panel) and the
entropy (bottom panel) measured in \RbFe\ in zero field.
                The dashed and solid lines correspond to the lattice and the magnetic contributions to the entropy.}
 \label{fig:CTip2}
\end{figure}

\begin{figure}
\includegraphics[width=0.95\columnwidth]{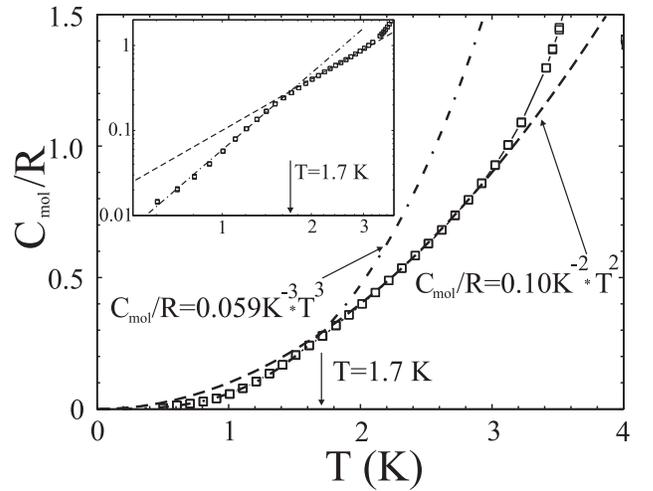}
\caption{Temperature dependence of the specific heat of \RbFe\ in the
low-temperature region.
               A crossover from a square to a cubic temperature dependence is clearly visible on the log-log plot (the inset).}
 \label{fig:CTip3}
\end{figure}

The top panel of Fig. \ref{fig:CTip2} shows the temperature dependence of
the specific heat measured in \RbFe\ for a wide temperature interval in
zero magnetic field. For temperatures in the region $10-40$~K, the
specific heat follows a $T^2$-type of the temperature dependence. As this
dependence is observed both below and above the Weiss temperature,
$\theta_{CW}=25$~K, it is reasonable to suggest that it corresponds
mostly to lattice vibrations. Extrapolation of the lattice contribution
down to zero temperature shows that for $T<5$~K, the magnetic
contribution to the specific heat is dominant and that the lattice
contribution can be neglected. The temperature dependence of the entropy
obtained by integrating the $C(T)/T vs \; T$ curves is shown in the
bottom panel of Fig. \ref{fig:CTip2}. By subtracting the lattice
contribution (the dashed line) one can obtain the temperature dependence
of the magnetic entropy, shown in Fig. \ref{fig:CTip2} as a solid line.
At temperatures exceeding several times the ordering temperature $T_N$
the magnetic entropy tends to a value agreeing well with ${\rm R}
ln(2S+1)={\rm R} ln6$.

Fig. \ref{fig:CTip3} shows temperature dependence of the zero-field
specific heat of \RbFe\ in the low-temperature region. For $T<1.7$~K the
temperature dependence of the specific heat is described well by the
regular cubic law (shown by the dash-dotted line in
Fig.~\ref{fig:CTip3}), while in the temperature interval 1.7 to 3.4~K it
follows a quadratic dependence (shown by the dashed line). The inset in
Fig.  \ref{fig:CTip3} reproduces the same data on a double logarithmic
scale.

\subsection{NMR \Rb\ spectra}
\begin{figure}
\includegraphics[width=0.95\columnwidth]{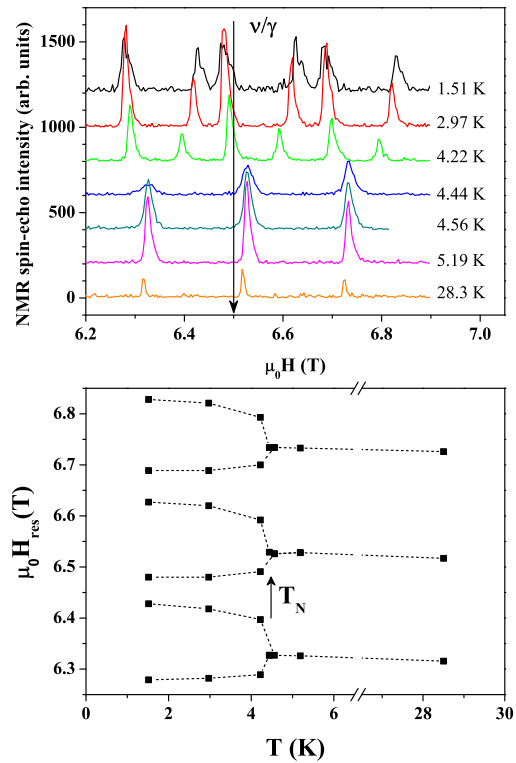}
\caption{(Color online)NMR spectra of \RbFe\ recorded for a frequency of
90.6~MHz at various temperatures, $\mathbf{H} \perp C_3$ (top panel).
                Bottom panel shows the temperature dependence of the resonance fields.}
\label{fig:90.6MHz(T)}
\end{figure}

The NMR \Rb\ spectra were recorded in the frequency range 35-115~MHz for
various temperatures. Typical temperature evolution of the spectra is
illustrated in Fig.~\ref{fig:90.6MHz(T)}, where the top panel shows the
field dependence of the spin-echo signal at different temperatures and
the bottom panel shows the temperature dependence of the resonance
fields. In the paramagnetic phase, the NMR spectrum consists of three
lines, with the central line corresponding to the ($-1/2 \leftrightarrow
1/2$) transition and the two satelites to the ($-3/2
\leftrightarrow-1/2$) and ($1/2 \leftrightarrow 3/2$) transitions. The
frequency difference for these transitions is caused by the quadrupole
splitting, which is found to be temperature independent over the entire
temperature range under investigation. The description and interpretation
of the NMR spectra in \RbFe\ and also their relation with the magnetic
structure are given in our earlier paper \cite{NMR}. As was shown in that
paper, the value of the NMR frequency in \RbFe\ is determined mostly by
the external magnetic field, by the quadrupole splitting and also by the
dipolar field created by the magnetic ${\rm Fe^{3+}}$ ions on \Rb\
nuclei, while the contribution from the hyperfine field is negligibly
small.

In the paramagnetic phase, on lowering the temperature, all three NMR
lines shift slightly towards higher fields and also broaden while
approaching the ordering temperature $T_N$. The shift is caused by the
dipolar field created on Rb sites by the field-induced magnetic moments
of the ${\rm Fe^{3+}}$ ions.

In the magnetically ordered phases, the dipolar fields of the Fe ions
make the Rb ion positions non-equivalent, which results in splitting of
all the spectral components into pairs of lines with distinct intensity.
The more intense line shifts towards lower field, while the less intense
line shifts towards higher field. The ratio of the intensities for these
lines is approximately 2 to 1. The observed splitting can be used for a
precise determination of the Neel temperature; it is also useful in
choosing an appropriate model for the magnetic structure of \RbFe
\cite{NMR}. The values of the ordering temperatures, as inferred from the
NMR results, are also collected in Fig. \ref{PhaseDiagr} together with
the magnetisation and specific heat points.

\begin{figure}
\includegraphics[width=0.95\columnwidth]{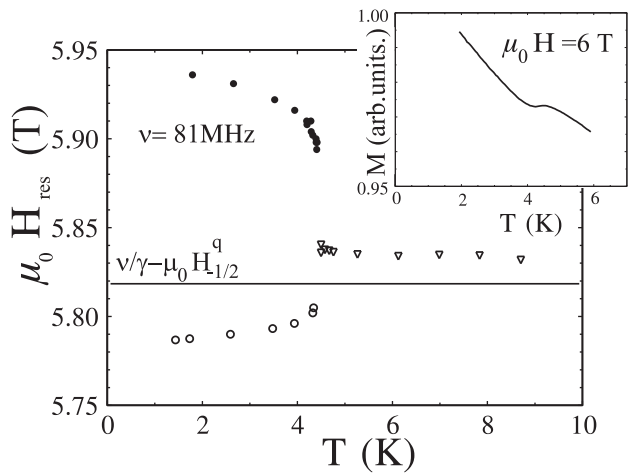}
\caption{Temperature dependence of the NMR resonance field $B_{res}(T)$
and the magnetic moment $M(T)$.
                $\mu_0 H \simeq 6$~T, ${\bf H} \perp C_3$.}
\label{NMR6T}
\end{figure}

In order to determine the temperature dependence of the order parameter,
detailed NMR measurements were performed near the critical temperature
for one of the spectral components. This component is seen at a frequency
of 81~MHz in a field of 5.82~T, which is nearly in the centre of the
collinear phase. The values of the resonance fields for a central line
above and below its splitting point are plotted in Fig. \ref{NMR6T}.

Let us define here the order parameter in the collinear phase following
Ref. \onlinecite{Miyashita2}:

${\bf \eta} = (({\bf M}_1+{\bf M}_2)/2)-{\bf M}_3)/2$, \\ where ${\bf
M}_1$, ${\bf M}_2$ and ${\bf M}_3$ are the magnetic moments of the three
sublattices normalised per magnetic ion. The parameter $\eta$ is zero in
the paramagnetic phase and nonzero in the ordered collinear phase. Let us
also introduce a normalised full magnetic moment:

${\bf \mu}=({\bf M}_1+{\bf M}_2+{\bf M}_3)/3$.

The temperature dependence of the order parameter $\eta$ can be
established by measuring the local field $B_{eff}$, which is created by
the Fe$^{3+}$ ions on the Rb nuclei. Experimentally, this field is
defined as

$B_{eff} = B_{res} - \nu / \gamma - \Delta \nu^q / \gamma$, \\ where
$\Delta \nu^q$ is the quadrupole shift of the resonance field.

At any finite temperature the magnetic structure can be represented as a
superposition of the ferromagnetic and collinear (UUD) states with the
corresponding weights $\mu - \eta/3$ and $\eta$, respectively. Therefore,
the effective field $B_{eff}$ consists of two components, one of them
being proportional to $\mu$, and the other being proportional to
$\eta(T)$. The order parameter is hence given by the expression

\begin{figure}
\includegraphics[width=0.95\columnwidth]{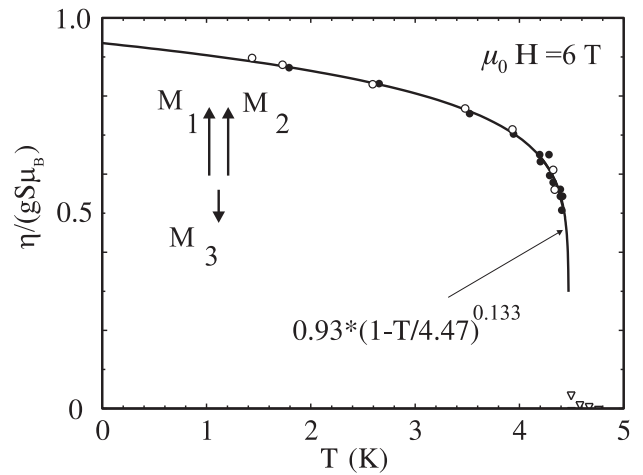}
\caption{Temperature dependence of the order parameter $\eta$.
               The solid and open circles correspond to the values of $\eta$ obtained from the resonance fields of \Rb\ residing in different positions with respect to the magnetic structure.
               The triangles correspond to the values of $\eta$ in the paramagnetic phase.
 $\mu_0 H \simeq 6$~T, ${\bf H} \perp C_3$.}
\label{OrderParameter}
\end{figure}

$\eta(T)\sim B_{eff}(T)-B_{eff}(T_0)\cdot M(T)/M(T_0)$, \\ where $T_0$ is
a fixed temperature above the ordering temperature $T_N$. From this
expression, by using the independently measured temperature dependence of
the magnetic moment, one can determine the order parameter.
Fig.~\ref{NMR6T} gives the temperature dependence of the NMR resonance
field as well as of the magnetic moment $M(T)$ measured in a field of
$\simeq 6$~T. Fig.  \ref{OrderParameter} shows the temperature dependence
of the order parameter obtained using the data of Fig. \ref{NMR6T} as
described above. An absolute value of the order parameter is obtained by
using the results of Ref. \onlinecite{NMR}, where the value of  $\eta$
was determined for $T = 1.6$~K and a magnetic field of 6~T. An abrupt
change in the order parameter is clearly seen near the ordering
temperature.

\subsection{Magnetic phase diagram}
\begin{figure}
\includegraphics[width=0.95\columnwidth]{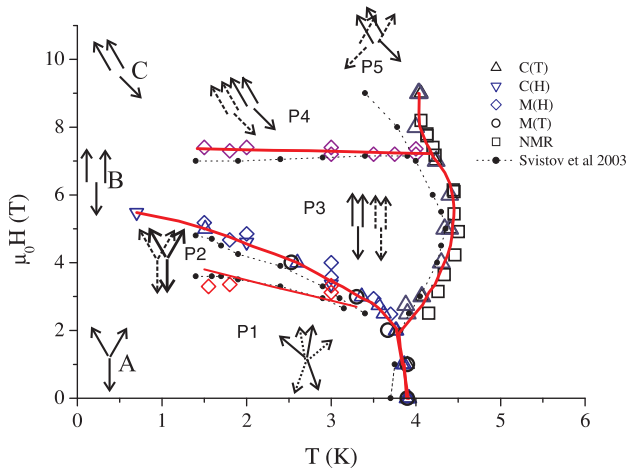}
\caption{(Color online)Magnetic phase diagram of \RbFe\ for $\mathbf{H}
\perp C_3$ determined by using different experimental methods. Spin
structures of magnetic layers are schematically presented by solid arrows
and marked as A, B, C. Two-layer fragments of supposed 3D spin structures
are shown by solid and dashed arrows for different magnetic phases marked
as P1-P5, see text. } \label{PhaseDiagr}
\end{figure}

Fig. \ref{PhaseDiagr} summarises the positions of the phase boundaries in
\RbFe\ for a field applied in the basal plane of the crystal, determined
by using different methods. In weak fields, the transition temperature is
well defined (with the accuracy better than 0.1~K) from the magnetisation
and specific heat measurements. The phase boundary between the
paramagnetic and the collinear phase (P3 in Fig. \ref{PhaseDiagr}) is
also well defined by the peak in the specific heat and by the splitting
in the NMR spectrum, while the magnetisation curve shows a smooth
behaviour. The phase boundaries corresponding to the transition fields
$H_{c2}$ and $H_{c3}$ are only visible in the magnetisation curves,
whereas the anomaly at $H_{c1}$ is clearly seen in both the specific heat
and the magnetisation curves.

The dashed lines in Fig. \ref{PhaseDiagr} represent the phase boundaries
obtained in our earlier paper \cite{Svistov}. Although the overall
agreement between the two sets of the results is good, a noticeable
difference between the transition temperatures is most likely to be
caused by the variations in sample quality, originating from difference
in the sample preparation procedures. The specific heat peaks associated
with the phase transition from the paramagnetic to the collinear phase
were much sharper for the new batch of samples, for which the NMR signal
splitting also occurs over a narrower temperature interval. During the
measurements with different samples from the same growth batch it has
been noticed that by rigidly gluing the NMR sample to a sample holder it
was possible to broaden significantly the phase transition region. In
addition, the specific heat measurements (performed as usual for the
heat-pulse method with the crystals attached rigidly to a platform using
a low-temperature grease) show that the peaks in $C(T)$ measured for the
PM to P3 transition in the thinner and therefore more stressed samples
were about 0.2~K broader compared to the thicker and less stressed
samples. It should be added, however, that the shape and the width of the
peaks in $C(T)$ curves were not sensitive to the stress for the other
magnetic phases (above and below the phase with a collinear structure).

\section{Magnetic phase diagram for $H \parallel C^3$}

\begin{figure}
\includegraphics[width=0.95\columnwidth]{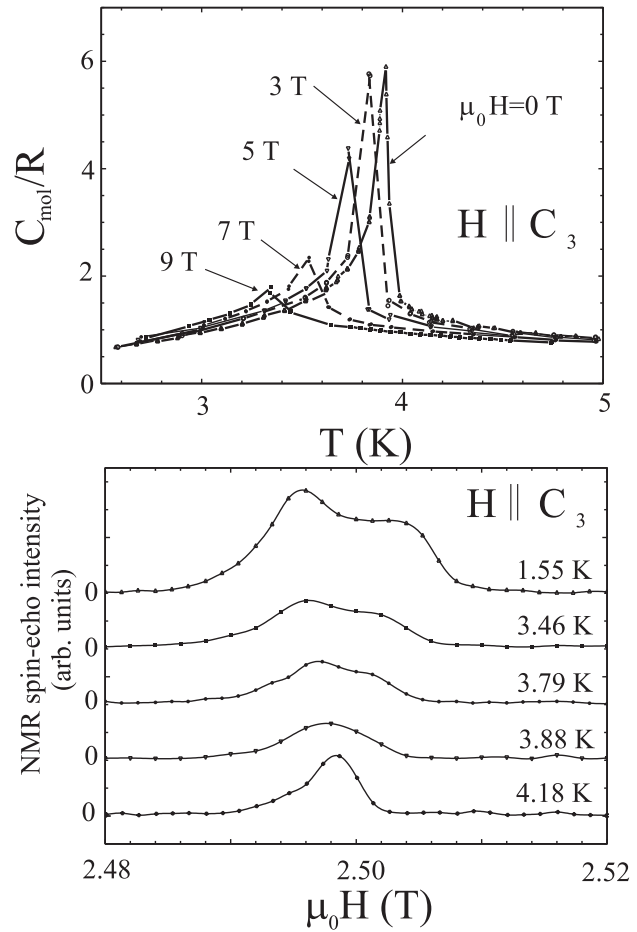}
\caption{Temperature dependence of the specific heat in different fields
(top panel) and spectra of the central quadrupole-split NMR line measured
at 35 MHz for different temperatures (bottom panel).
                The field is applied along the $C^3$ axis of the \RbFe\ single crystal.}
\label{CToup}
\end{figure}

Fig. \ref{CToup} (upper panel) shows the temperature dependence of the
specific heat measured in \RbFe\ for ${\bf H} \parallel C^3$. The lower
panel also shows the records of the central quadrupolar-split NMR line
measured at 35~MHz for the same direction of the field. When the field is
applied along the $C^3$-axis (``hard" axis in terms of the magnetic
anisotropy), a single phase transition into a magnetically ordered phase
is observed via a $\lambda$-like anomaly in $C(T)$. This transition is
seen as a sharp maximum in the $C(H)$ curves (not shown) and also as a
splitting of the NMR line. For ${\bf H} \parallel C^3$, this splitting is
nearly an order of magnitude smaller compared to the ${\bf H} \perp C^3$
geometry.

\begin{figure}
\includegraphics[width=0.95\columnwidth]{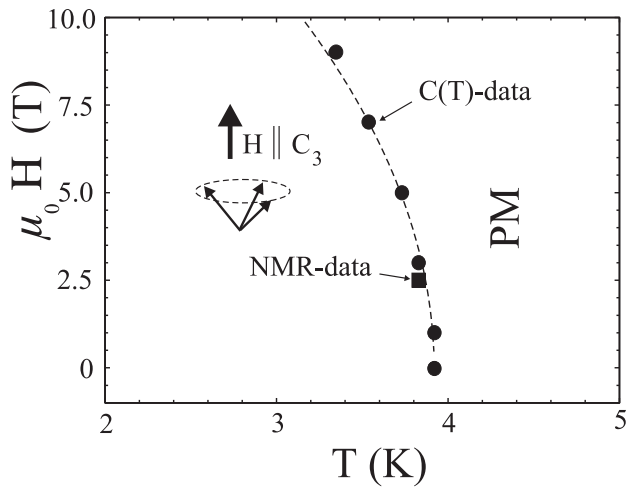}
\caption{The magnetic phase diagram of the \RbFe\ single crystal for
$\mathbf{H} \parallel C^3$.
                The dashed line corresponds to the relation
                $\mu_0$H $= 20.2$ T $ \times (1-T/3.9$ K$)^{0.5}$.
                The value of the experimental uncertainty in the position
                of the phase boundary is represented by the markers size.}
 \label{PDoupl}
\end{figure}

The $H-T$ phase diagram for a field applied along the $C^3$-axis is shown
in Fig. \ref{PDoupl}. The circles correspond to the phase boundary
positions obtained from the specific heat measurements, while the square
marks the phase transition observed by the NMR splitting.

\section{Discussion}
\subsection{Initial aspects}

Here we discuss the observed phase transitions in \RbFe\ and its critical
properties. Apart from the data presented in this paper, the discussion
will be based on the available data on the microscopic structure of \RbFe\
obtained from neutron diffraction \cite{Broholm2002} and NMR \cite{NMR}
experiments, as well as on the established hierarchy of the magnetic
interactions \cite{Svistov}:  the intralayer exchange, the easy-plane
anisotropy and the interlayer exchange. Besides these experimental facts
we shall use the results of the calculation of the dipole energy, which
can help to select between the proposed states with equal  nearest
neighbors  exchange energy.

 It has to be noted that in zero field and at finite temperature for
the case of an ideal two-dimensional system, long range order of the
magnetic components is prohibited by the Mermin-Wagner theorem. For a
triangular lattice with \afm ic exchange interactions, however, apart from
the Berezinskii-Kosterlitz-Thouless transition there should be an
additional transition to a state with long-range order with the chirality
parameter for the neighbouring triangles, the so-called ``staggered
helicity (vorticity)" state \cite{Miyashita,Korshunov}. Monte Carlo
simulations for this state \cite{Miyashita} show that this phase is close
to the state with long-range order of the nonzero average magnetic
moments, as the spin-spin correlations decay in a power-law manner.
Moreover, the characteristic relaxation time for the 120-degree
three-sublattice structure could be relatively long and may exceed the
dynamical range for various experimental techniques.

Computer simulations show the presence of chiral ordering both for the 2D
XY-model \cite{Landau} and 2D Heisenberg
system.\cite{Miyashita,Miyashita2} For the 2D XY-model, in addition to
the paramagnetic phase, the $H-T$ phase diagram contains the three
ordered phases,\cite{Landau} which have previously been identified in,
for example, Ref. \onlinecite{Korshunov}. These structures are depicted
and labelled as A, B and C in Fig. \ref{PhaseDiagr}.

\begin{figure}
\includegraphics[width=0.95\columnwidth]{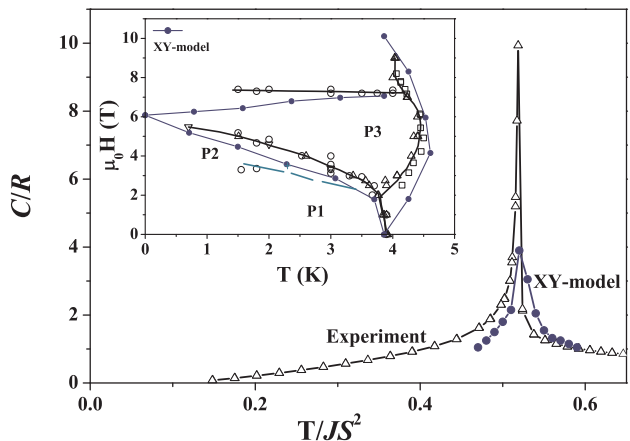}
\caption{(Color online) Experimental (open symbols) and theoretical
\cite{Landau} (solid symbols) temperature dependence of the specific heat
in \RbFe\ plotted against the normalised temperature in zero field.
               The inset shows the $H-T$ phase diagram for $\bf{H} \perp C_3$,
                where the open symbols represent the experimental
                results as in Fig.~\ref{PhaseDiagr}, solid symbols are
                the results of computer simulations \cite{Landau} with
                $J = 1.23$~K.
                The solid lines in the inset are drawn as a guide to the eye.}
\label{PDLandau}
\end{figure}

In the presence of an external magnetic field, the stabilisation of long
range order becomes inevitable \cite{Korshunov}. Computer simulations for
the Heisenberg spins \cite{Miyashita,Miyashita2} reveal qualitatively
similar results to the XY-model phase diagram, however, the exact
positions of the phase boundaries are determined there far less
accurately. Both models predict a characteristic non-monotonic variation
of the ordering temperature with the magnetic field. The simulation
results for the XY-model are shown in Fig. \ref{PDLandau}.

Weak interlayer exchange coupling should, in the general case, facilitate
the appearance of 3D long range order. Given the results for the 2D
models discussed above and the proximity of the ground states for these
systems, one can speculate that the temperature of the 3D ordering should
be close to the critical temperature of the 2D transition. In order to
corroborate this conjecture, additional Monte Carlo simulations for the
system with a finite interlayer exchange and magnetic anisotropy would be
required.

In order to analyse the real structures in 3D crystals one has to take
into account the interlayer exchange coupling and to consider various
possibilities for the alignment of the spins in the neighbouring planes.
Theoretical analysis of the different allowed structures for an \afm ic
interlayer exchange has been performed in Ref.~\onlinecite{Gekht}. The
proposed magnetic phases have then been used as the models for \RbFe\ in
Ref.~\onlinecite{Svistov}. Fig.~\ref{PhaseDiagr} shows the spin
configurations of the ordered phases, which have been selected from the
proposed list \cite{Gekht} on the basis of matching the NMR-detected
commensurate-incommensurate transition \cite{NMR}, magnetisation
plateau, and possible periodicity along  $C^3$ direction, which is
discussed below. Solid and dashed arrows in Fig.~\ref{PhaseDiagr}
represent the orientations of the magnetic moments of the Fe$^{3+}$ ions
from neighboring planes for these selected configurations.

\begin{figure}
\includegraphics[width=0.95\columnwidth]{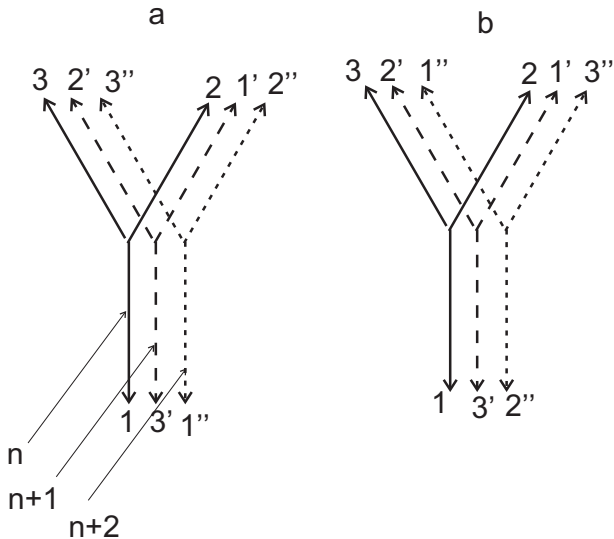}
\caption{Schematic representation of the magnetic structures with
different periodicity along the $C^3$-axis: $2c$ on the left (a) and $3c$
on the right (b).
                Solid and dashed arrows correspond to the direction of the magnetic moments in the neighbouring planes. Numbers 1, 1' and 1'' mark the magnetic moments of the Fe$^{3+}$ ions located above one another in the neighbouring  $n^{th}$,  $n+1^{st}$ and  $n+2^{nd}$ magnetic layers.}
\label{Fig_ADD}
\end{figure}

Let us consider the procedure of the selection of the proposed structures.
One should note that for the spin structures with two parallel sublattice
magnetisations within a layer (e.g. structures suggested for phases
P2,P3,P4), an interlayer exchange does not define the magnetic periodicity
along the $C^3$-axis, which could therefore be any integer multiple of the
lattice periods. This point is illustrated in Fig.~\ref{Fig_ADD}, where
the two structures with the periodicity along the $C^3$-axis equal $2c$
and $3c$, contain the same number of various angle combinations for the
neighbouring spins. These two structures thus possess the same exchange
energy. This periodicity must be due to some other interactions, that are
weaker when compared to the exchange coupling. Among the possible
candidates for the interactions which select a period along $C^3$ axis are
the dipole-dipole interactions and further neighbour exchange interactions
along the $C^3$-axis. The direct calculation of the dipole energy shows
that it is lower for the period of $3c$. This gain in the dipole energy
equals approximately 0.4 mK per Fe-ion for UUD structure. Therefore we
suppose that the structures suggested for phases P2, P3, P4 have the
period $3c$. The next-nearest antiferromagnetic exchange interaction in
the $C^3$ direction would also result in the preference of period $3c$.

Fig.~\ref{PhaseDiagr} shows results of the various experimental techniques
that reveal the presence of five ordered phases. The phase transition,
corresponding to the field $H_{c3}$, which is marked in the
$\frac{dM}{dH}$ curves in Fig.~\ref{fig:MvsH} is an
incommensurate-commensurate transition indicated by the characteristic
change of the NMR spectrum \cite{NMR}. We suppose that the incommensurate
structure below $H_{c3}$ corresponds, as usual, to a nearly antiparallel
orientation of spins from neighboring layers (structure $P1$), this
structure being rotated around the $C^3$ axis by a small angle from layer
to layer. In the range above $H_{c3}$ the commensurate structure takes
place for the phase P2. In Ref. \onlinecite{Svistov}, two possible
variants were proposed for the spin structures in phases $P1$ and $P2$
based on the results of theoretical considerations
\cite{Gekht,Korshunov2002}. We selected  for the phase $P2$, the structure
which allows the period of $3c$, favorable for dipole-dipole energy. For
the phase $P1$ the structure, compatible with period $2c$ is suggested,
since for the typical incommensurate structures in antiferromagnets the
orientation of spins in neighboring layers should be close to antiparallel
one, like at the period value $2c$.

In the phase $P3$, i.e. in the plateau range, the collinear structure
within a layer and the periodic structure along $C^3$-axis, probably
with the period $3c$, is realised.
 The structures of P4 or P5 type may be proposed for the high-field range
between $H_{c2}$ and saturation. We have here no NMR data which could
distinguish between commensurate and incommensurate states, thus we can
not determine the nature of the P4-P5 boundary, detected by small $dM/dH$
singularity in Ref\cite{Svistov}. Nevertheless, if we extrapolate  the
period value of $3c$ from $P3$ to $P4$ , the structure which is compatible
with period $3c$ should be selected, as shown in Fig.~\ref{PhaseDiagr}. In
this case the transition to the second high-field spin structure of the
phase $P5$ should change the period to $2c$ or again result in the
incommensurate modulation.

The collinear structure of the $P3$ phase may be taken as determined
most precisely, as it corresponds to the maximum value of the NMR shift
\cite{NMR}, which is maintained in the entire field range
$H_{c1}<H<H_{c2}$. The spin structures of other phases are suggested as
probable configurations.

\subsection{2D-3D crossover of the specific heat temperature dependence}
The overall shape of the phase diagram, as well as the characteristic
widening of the collinear phase range with increasing temperature
predicted in Ref.~\onlinecite{Korshunov} points to the two-dimensional
nature of magnetism in \RbFe. Here, we show that the observed change-over
in the temperature dependence of the specific heat from the
low-temperature $T^3$ law to the intermediate temperature $T^2$ law (as
described in section \ref{subsec_Magnet}) is also indicative of the
two-dimensional disposition of this magnetic system.

The temperature dependence of an \afm\ for $T < 0.6 \; T_N$ is usually
determined by thermally activated magnons. The spectrum of acoustic
magnons in \afm ic \RbFe\ consists of one gapless mode and two modes with
a gap of 90~GHz (4.5~K).\cite{Svistov}  At such low temperatures, the
variation of the gap's magnitude should be small, as the order parameter
for $T < 0.6 \; T_N$ is reduced by not more than 30\%. The main
contribution to the specific heat in this temperature region comes from
the magnons of the gapless branch. At the lowest temperature the
wave-numbers of the thermal magnons are small and lie far away from the
Brillouin zone boundary. Therefore one can expect for $C(T)$ a typical
three-dimensional behaviour, that is $C(T) \sim T^3$.

Due to a large difference between the inter- and the intra-layer exchange
interactions, the magnon dispersion relation is highly anisotropic -- the
magnons propagating along the $C^3$-axis are less dispersive than those
with a wave-vector confined to the triangular plane. Consequently, at a
certain temperature, the energy of the thermal magnons travelling along
the $C^3$-axis reaches its maximum value, while the size of the $k$-space
region corresponding to magnons with $k \perp C^3$ will continue to
expand. For these temperatures one can expect a typical two-dimensional
behaviour, that is $C(T) \sim T^2$. Inevitably, the change-over from a
cubic to a quadratic law should occur at the temperature dependent on
interlayer exchange energy, or, more precisely, on the maximum energy of
the $k || C^3$ magnons of the lowest gapless branch. This is the energy of
the lowest ``exchange"-mode of the antiferromagnetic resonance spectrum.
According to Ref. \onlinecite{Svistov}, the gap of the lowest exchange
mode  is about 30~GHz (1.5 K), which agrees well with the observed
crossover temperature. Thus the 2D-3D crossover in the temperature
dependence of specific heat confirms the domination of two-dimensional
features in the thermodynamic behaviour of \RbFe.

\subsection{Critical properties of the magnetic ordering transition
for ${\bf H} \perp C^3$}

Given the two-dimensional nature of the magnetic properties of \RbFe, let
us compare the observed critical behaviour with the results of the
classical Monte Carlo simulations \cite{Landau} performed for the 2D
XY-model. The Neel temperature in this simulation is defined as
$0.5JS^2$. Using the value of $J=1.2 \pm 0.1$~K determined from the
susceptibility and saturation magnetisation measurements
\cite{Svistov,Inami}, the predicted ordering temperature is $3.75 \pm
0.3$~K, which is in a good agreement with the value of $T_N=3.85 \pm
0.05$~K determined experimentally. The ordering temperature for the 2D
Heisenberg \afm\ on a triangular lattice, on the other hand, is defined
as $0.35JS^2$, \cite{Miyashita2} which unexpectedly gives less
satisfactory agreement with the experimental $T_N$.

Fig. \ref{PDLandau} shows the comparison between the temperature
dependence of the specific heat measured experimentally and obtained
from the Monte Carlo simulations \cite{Landau}. It also shows a
comparison of the phase boundary positions from the experimental $H-T$
diagram (see Fig. \ref{PhaseDiagr}) and the simulations \cite{Landau} for
$\bf{H} \perp C^3$. For the theoretical curve, the value of the exchange
parameter quoted above has been used. The overall agreement is
satisfactory, as the two phase diagrams have very similar shapes, apart
from the low-field region. For low fields, however, the distinction is
fundamental: the theoretical model predicts a phase transition from
paramagnetic to chiral structures only through a collinear phase, while
experimentally a single phase transition is observed for all fields below
2~T.

\begin{figure}
\includegraphics[width=0.95\columnwidth]{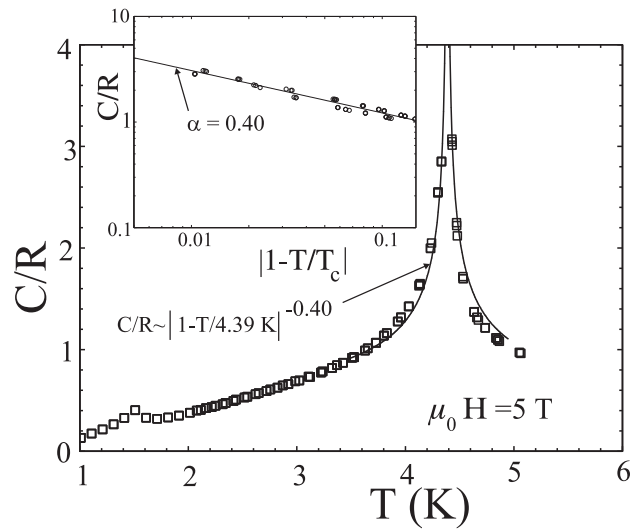}
\caption{Temperature dependence of the specific heat for a field of
$\mu_0 H = 5$~T applied in the easy-plane of \RbFe.
 The inset shows $C \; vs \; |1-T/T_c|$ curve in the vicinity of the
 phase transition on a log-log scale.}
 \label{C5T}
\end{figure}

The measured $C(T)$ and $\eta(T)$ curves allow one to determine the
specific heat and the order parameter critical indices for the phase
transition into the collinear phase. Corresponding experimental data are
presented in Figs. \ref{OrderParameter} and \ref{C5T}. The experimentally
determined indices are $\alpha = 0.40 \pm 0.03$ for the specific heat and
$\beta = 0.13 \pm 0.02$ for the order parameter. They agree well with
$\alpha=1/3$ and $\beta=1/9$ obtained from the simulations \cite{Landau}.

It can be concluded then that the ordering temperature, the magnetic
phase diagram and the critical indices of the specific heat and the order
parameter demonstrate a satisfactory quantitative agreement with the 2D
XY-model predictions for zero field and also for regions of the field ($H
\perp C^3$) where the collinear phase is stable. Most likely this
agreement can be explained by the large value of the energy gap for the
magnon branch, which corresponds to a deviation of the spins out of the
basal plane.

This gap has been estimated as 90 GHz (or 4.5 K) \cite{Svistov}.
Therefore, for all the magnetically ordered phases, the thermodynamic
properties are defined by gapless and low-energy magnons, corresponding
to spin fluctuations within the easy plane. This, in turn, justifies the
applicability of the XY model.

\subsection{Phase diagram for $\bf H \parallel C^3$}
The easy-plane type of magnetic anisotropy in \RbFe\ is confirmed by the
observation of the magnetisation plateau around $M_s/3$ for the magnetic
field applied in the $ab$ plane and by the linear rise of the $M(H)$
curves for the field parallel to the $C^3$-axis. For an easy-plane
anisotropy, the magnetic moment of the system is generated for $H
\parallel C^3$ by the even tilts of all the subblattices in an
umbrella-like structure towards the field (see Fig. \ref{PDoupl}). For
such a system, the application of an external magnetic field should
result in a decrease of the ordering temperature. Moreover, in the
vicinity of $T_N(H=0)$, the reduction should be proportional to the
square of the field, similarly to an ordinary two-sublattice \afm\
\cite{LandauLifshitz}. The dashed line in Fig. \ref{PDoupl} shows such a
dependence, $H_c=a(1-T/T_c)^{0.5}$, with $a=20.2$~T and $T_c=3.9$~K. For
the case of the uniformly tilted spins described here, splitting in the
NMR spectra should be absent, but it is observed. The origin of this
small but clearly visible splitting remains unknown at the present time.

\section{Conclusions}
To summarise, we have performed a systematic and comprehensive study of
the thermodynamic and magnetic properties of the \qtd\ TLAM \RbFe, as
well as measurements of the magnetic order parameter for this compound. A
crossover from a $T^2$ law to a  $T^3$ law is observed for the
temperature dependence of the specific heat at low temperatures. The
phase diagram and critical properties demonstrate a surprisingly good
quantitative agreement with the predictions for the XY-model despite the
fact that the easy-plane anisotropy in \RbFe\ does not dominate the main
exchange interaction.

The authors are grateful to S.E.~Korshunov, V.I.~Marchenko,
M.E.~Zhitomirsky, A.~Loidl, H.-A.~Krug~von~Nidda for discussions and to
M.R.~Lees for a critical reading of the manuscript. This work is
supported by the Grant  04-02-17294 Russian Foundation for Basic
Research, RF President Science Schools Program, contract BMBF \textnumero
VDI/EKM 13N6917-A, program of the German research society
Sonderforschungsbereich 484 (Augsburg), and the EPSRC grant (University
of Warwick). The work of L.E.~Svistov is supported by the Alexander von
Humboldt fellowship.

\end{document}